      [1999/12/01 v1.4c Il Nuovo Cimento]
\documentclass{cimento}
\usepackage{amsmath}

\title{High performance computing for classic gravitational N-body systems}
\author{R.~Capuzzo-Dolcetta}
\instlist{\inst{} Dept. of Physics, Sapienza, Universit\' di Roma, Roma, 
Italy}
\PACSes{\PACSit{02.70.Ns}{Molecular dynamics and particle methods}
\PACSit{95.10.Ce}{Celestial mechanics (including n-body problems)}
\PACSit{98.10.+z}{Stellar dynamics and kinematics}}

\begin{document}

\maketitle

\begin{abstract}
The role of gravity is crucial in astrophysics. It determines the evolution of
any system, over an enormous range of time and space scales. Astronomical stellar systems as composed by N interacting bodies represent examples of self-gravitating systems, usually treatable with the aid of newtonian gravity but for particular cases. 
In this note I will briefly discuss some of the open problems in the dynamical study of classic self-gravitating N-body systems, over the astronomical range of N.  
I will also point out how modern research in this field compulsorily
requires a heavy use of large scale computations, due to the contemporary
requirement of high precision and high computational speed.
\end{abstract}

\section{Introduction}
In terrestrial physics gravity is, of course, important but not
difficult to account for because it simply corresponds to an external constant
field to add to other more complicated interaction among the constituents of the
system under study. The physical systems on earth are not
self-gravitating, and this corresponds to an enormous simplification. In an
astrophysical context, it is no more so.
Astronomical objects are self-gravitating; their shape, volume and dynamics are
determined mainly by self-gravity. It acts, often, in conjunction with the external
gravity due to the presence of other bodies, which makes the object under
study moving on some orbit, and influences its shape, at least in its outskirts,
by mean of tidal interactions.

A simple parameter to measure how much self-gravity contributes to the whole
energetics of a given system may be the ratio, $\alpha$, between the self-gravitation
energy of the system and the energy given by the external gravitation field
where the system is embedded in.
For a typical terrestrial system like the Garda lake 
$\alpha \simeq 10^{-8}$, while for
two astronomical systems (a typical globular cluster moving in a galaxy and a typical galaxy in a
galaxy cluster) $\alpha \simeq 10^{-2}$: a million times greater. A part
from the other, obvious, differences (a lake is composed by a liquid, where the
collisional time scale is negligible respect to any other time scale in the
system, while the globular cluster and the galaxy are composed by stars moving in volumes 
such that the collisional 2-body time scale is comparable,
in the case of globular cluster, or much longer, in the case of galaxy, to the
system orbital time and age), it is clear that while the lake molecules mutual
gravitational interactions are negligible respect to the external field, this is not the case for the stars in
globular clusters or galaxies.

\section{Astronomical N-body systems}
As stated in the Introduction, self-gravity cannot be neglected when dealing
with
physics of astronomical objects. This makes theoretical astrophysics a hard
field: astrophysical systems are intrinsecally difficult to study, even in
newtonian approximation,
because of the {\it double divergence} of the, simple, two-body interaction
potential, $U_{ij}\propto 1/r_{ij}$, where $r_{ij}$ is the euclidean distance between the $i$ and $j$ 
particle,\\
$r_{ij}=\sqrt{(x_i-x_j)^2+(y_i-y_j)^2+(z_i-z_j)^2}$. {\it Ultra-violet} divergence corresponds to very close
encounters, {\it infra-red} divergence to that the gravitational interaction never vanishes. 
These divergences introduce a multiplicity of time scales (\cite{ref:aar}) and make impossible to rely on statistical mechanics and/or to
non-perturbative methods, as often done in other particle-systems physics.
Actually, the newtonian N-body dynamics is mathematically represented by the
system of N second-order differential equations

\begin{equation}
\label{system}
\left \{ \begin{array}{ll}
\ddot{\textbf{r}_i}=\sum_{\substack{j=1\\j\neq i}}^
{N}G\frac{m_{j}}{\textbf{r}_{ij}^{3}}\,
(\textbf{r}_{j}-\textbf{r}_{i}),\\
\dot{\textbf{r}_i}(0)=\dot{\textbf{r}_i}_{0},\\
\textbf{r}_i(0)=\textbf{r}_{i0},\\
(i=1,2, \dots ,N).
\end{array}
\right.
\end{equation}

\noindent This dynamical system is characterized by: (i) $O(N^2)$ complexity, (ii) being far from linearity, (iii) having few constraints in the phase-space. Sundman (\cite{ref:sun}) in 1912 showed (and won the King Oscar II Prize) that there exists a series solution in powers
of $t^{1/3}$ convergent for all $t$, except initial data
which correspond to zero angular momentum. This result was generalized to any N
only in 1991 by \cite{ref:wan}. 
Anyway, the power series solutions are so slow in convergence
to be useless for practical use. This means that the gravitational N-body
problem must be attacked numerically.
The difficulties in doing this are, contemporarily, {\it theoretical} and {\it
practical}. On the {\it theoretical} point of view, one has to face with the chaotic
behaviour of the nonlinear system which is related to the extreme sensitivity
of the system's differential equations to the initial conditions: a
very small initial difference may result in an enormous change in the 
long-term behaviour of the system. Celestial dynamics gives, indeed, one of the
oldest examples of chaos in physics. This problem is almost
unsolvable; it may be just kept under some control by using sophisticated, high
order time integration algorithms. On the {\it practical} side, the (obvious) greatest
complication to face is the due to the infrared (large scale) divergence, that
implies the need of computing all the $\propto N^2$ force interactions between
the pairs in the systems. This results in an extremely demanding computational
task, when N is large (see Table I).
We will now discuss some of the problems arising when dealing with the
numerical study of the evolution of self-gravitating systems over the
astronomical range of N.

\begin{table}[]
\begin{center}
\caption{Some typical astronomical systems, with their star number (N), 
number of floating point operations needed for the force evaluations 
in a single system configuration ($n_f$) and CPU time  
required to the $n_f$ operations  by a single processor of 1 Gflops speed (t$_{CPU}$, in seconds).
Note that $1.8\times 10^{14}$ sec $\simeq 5.7$ Myr!}
\label{cputimes}
\begin{tabular}{|l|c|c|c|} \hline \hline
system  & N    & n$_f$ & t$_{CPU}$\\ \hline
Open cluster & $1000$  & $1.5\times 10^7$ & $0.02$  \\ \hline
Globular cluster & $10^5$  & $1.5\times 10^{11}$ & $180$\\ \hline
Galaxy & $10^{11}$ & $1.5\times  10^{23}$ & $1.8\times  10^{14}$ \\ \hline
\hline
\end{tabular}
\end{center}
\end{table}

\section{Small- and Large- N systems}
On the small- N side ((N $\leq 10$, example: solar system) the problem is not
that of enormous CPU time consumption, for the number of pairs is small, but,
rather, that of the need of an enormous precision. This to keep the round-off error within 
acceptable bounds when integrating over many
orbital times. In the case of few bodies, reliable investigations cannot accept
the point mass scheme (for instance, the Sun potential requires a multipole
expansion) and high precision codes are compulsory. Pair force evaluation is
computationally cheap due to the low number of pairs; 
on the other side, even very small round-off errors increase secularly, time step by time step, making high order
symplectic integration algorithms unavoidable. The need is: a fast computer,
able to handle with motion integration over a very extended time and able to
evaluate forces with enormous precision.\\ 
We do not speak any further of the few body regime, which is the realm of modern
celestial mechanics and space dynamics, but go to say something on the problem of intermediate- and large-N-body systems, task which
is typical of the modern stellar dynamics, instead.
Force computation by pairs is computationally expensive, the mostly demanding part being 
the evaluation of the distance $r_{ij}$ between the generic $i$ and $j$ particle. It requires the computation of a square root which, still with modern computers, is based on
ancient methods among which the Erone's method, the Bombelli's method and the Newton-Raphson numerical solution of the quadratic equation $x^2- r_{ij}^2=0$. In any case, the single pair
force evaluation requires about 30 floating point operations; this means that 
in an N-body system, $n_f = 30\times N(N-1)/2$ floating point operations 
are required. 
A single processor (PE) with a speed of 1 Gflops would compute the single pair
force in $\sim 3\times 10^{-8}$ sec. Consequently, the whole N star forces
would require the time indicated in Table I for their evaluation at every time
step. Clearly, the task of following numerically the long term evolution of a large-
N-body system by a program based on direct summation of pair forces is very far
out of the capability even of the most performing computers.
Actually, the profiling of any computer code to integrate N-body evolution
indicates that about $70\%$ of the CPU time is spent in force evaluation. 

What strategies must be used, then?

The most natural way to attack the problem is a proper combination of the
following ingredients: 
(i) simplification of the interaction force calculation; (ii) reduction of the
number of times that the forces have to be evaluated, by a proper
variation of the time step both in space and in time; (iii) use the most
powerful (parallel) computers available. Points (i) and (ii) require high level
numerical analysis, point (iii) requires the solution of the, not easy, problem
of parallelizing an N-body code. 

The simplification of force calculation may be done by mean of the introduction
of space grids, both for computing the large scale component of the
gravitational force via the solution of the Poisson's equation (with 
Fast--Fourier codes, for example) and for the dynamic subdivision of the space domain with 
a recursive, octal tree to take computational advantage by a multipole 
expansion of the interaction potential (approach first used by \cite{ref:bhu}). 
These are two of the possibilities to reduce the particle-particle (PP) force evaluation 
to a particle-mesh (PM) or particle-particle-particle-mesh (P3M) approach, 
with obvious computational advantages (see \cite{ref:hoe} for a general discussion). 
In addition to the complications introduced in the computer code, a clear limit of this procedure is the error introduced in the force evaluation,
which can be reduced, over the small scale, by keeping a direct PP force
evaluation for close neighbours. Point (ii), time stepping variation, relies mainly
on the use of individual (per particle) time steps.
Particles are advanced with a time step  proper to the
individual acceleration felt, allowing a reduction in highly
dynamical cases without stopping the overall calculation. Unfortunately,
individual time stepping requires 
careful implementation to guarantee synchronous integration and implies, often,
a reduction of order of precision of the integration method. Finally, the
parallelization of gravitational codes (point (iii)) is difficult, because
gravity is such that the force on every particle depends on the position of all
the others. This makes non trivial a domain decomposition such to release a 
balanced computational weight to the various PEs of a
parallel machine. In this context, it is relevant noting that many active groups
of research chose to use \lq dedicated\rq parallel architectures, which act as boosters
of specific computations, like those of the distances between
particles. This is the road opened by the Japanese GRAPE group, lead by Makino (see \cite{ref:mak}).
Another, intriguing, possibility is the use of Graphic Processing Units (GPUs)
as cheap alternatives to dedicated systems.
GPUs are used to speed up force computations and give high computing performances at 
much lower costs, especially in cases where double precision is not required.
This is the choice explored in astrophysics first by 
S. Portegies Zwart and his dutch group (\cite{ref:spz}).  
Capuzzo-Dolcetta and collaborators in Italy have implemented a direct N-body code using as force evaluation accelerator
the brand new NVIDIA TESLA C1060 GPU and with a sophisticated 6$^{th}$ order 
symplectic integration performed by the host (\cite{ref:cdm}).

\end{document}